\documentclass[11pt]{revtex4}
 \pdfoutput=1 
\usepackage{graphicx}

\begin{document}
\title{Derivation of an equation of pair correlation function from BBGKY hierarchy in a weakly coupled self gravitating system}
\author{Anirban Bose}
\affiliation{ Serampore College, Serampore, Hooghly 712201, India}
%\date{\today}
%\maketitle
%\tableofcontents
\begin{abstract}An equation of pair correlation function has been derived from the first two members of BBGKY hierarchy in a weakly coupled inhomogeneous self gravitating system in quasi thermal equilibrium. This work may be useful to study the thermodynamic properties of the central region of a star cluster which is older than a  few or more central relaxation time.
\end{abstract}
%%%%%%%%%%%%%%%%%%%%%%%%%%%%%%%%%%%%%%%%%%%%%%%%%%%%%%%%%%%%%%%%%%%%%%%%%%%%%%%%%%%%%%%%%%%%%%%%%%%%%%%%%%%%%%%%%%%%%%%
\maketitle
\newpage
\section{Introduction}

Finite, bound self gravitating system has negative heat capacity.
The thermodynamics \cite{kn:spi,kn:sc,kn:ph1,kn:jk, kn:tt,kn:ac,kn:fb2,kn:fb3} of such system differs significantly from that of normal laboratory system and there is a question that if self gravitating system, where the constituent particles are interacting with each other via long range gravitational force, may at all attain the thermal equilibrium state.

In fact, absolute thermal equilibrium is not possible.  Since, in contrast to the laboratory system, there is no physical  container to confine the system to a particular place. Therefore, evaporation  of particles from the high energy tail of the distribution function does not allow the system to reach thermal equilibrium. 
 
This confinement problem may be theoretically resolved by the introduction of artificial boundary. Thermodynamic properties of such system have been extensively investigated \cite{kn:va,kn:dl,kn:tp} by confining N point particles each with mass m in a  spherical container.  The system is assumed to be thermally conducting. Therefore, the equilibrium state of this system is expected to be the isothermal sphere.
But that is not the actual fact. For both canonical and microcanonical ensembles instabilities occur if the density contrast between the centre and edge of an isothermal gas exceeds certain values. For microcanonical ensembles this phenomenon is recognised as gravothermal catastrophe \cite{kn:va,kn:dl,kn:tp} and such system can increase entropy by settling down to a state which is not isothermal.

However, self gravitating system like globular cluster whose relaxation time is less than the age of the cluster has a chance to reach close to the thermal equilibrium state. At least, the central part of a cluster older than a few central relaxation time is close to thermal equilibrium\cite{kn:sc}. These globular clusters are approximately spherical in shape and may be modelled by the N particle system confined in a  spherical container.

There are different theoretical approaches \cite{kn:spi,kn:sc,kn:hey,kn:ph3,kn:jb,kn:ham}to study the self gravitating system. Probably, the most common way is to explore the  Fokker-Planck equation \cite{kn:spi,kn:sc} of the concerned system. 

There are other ways.
For example, integrating the N particle Liouville equation over the phase space of (N-1) particles we obtain  the first member of BBGKY hierarchy equation which, in the limit of $N \rightarrow \infty$, becomes the collisionless Vlasov equation. This equation deals successfully with the self gravitating system where the influence of encounters are neglected. Integrating the Liouville equation once again over the phase space of (N-2) particles we obtain the second member of BBGKY hierarchy. These two equations, in the limit of large but finite values of N, are capable of exploring the thermodynamics of the system with weak encounters. \cite{kn:hag,kn:hey}.

In this article we attempt to derive an equation of pair correlation function from the first two members of BBGKY hierarchy for the inhomogeneous self gravitating system in quasi thermal equilibrium. This method has been previously applied\cite{kn:ab} to study the thermodynamic properties of the weakly correlated inhomogeneous plasma system. 
\section{Derivation of the equation of pair correlation in thermal equilibrium}

We consider a  self gravitating system of N particles enclosed in a spherical container. 
We assume, for simplicity, that all the particles have same mass m. This system is older than few collisional relaxation time so that the distribution function is in thermal equilibrium. Under these assumptions, the first two members of BBGKY hierarchy are written as : 

\begin{equation} \frac{\partial f_1}{\partial t} + {\bf v_1}\cdot\frac{\partial f_1}{\partial {\bf x_1}}
+ n_0\int d{\bf X_2} {\bf a_{12}}\cdot\frac{\partial}{\partial{\bf
v_1}}\left [ f_1({\bf X_1})f_1({\bf X_2})+g_{12}\right ] = 0
\label{v1}\end{equation}
 \begin{eqnarray}
&&\frac{\partial g_{12}}{\partial t } + {\bf v_{1}}\cdot
\frac{\partial g_{12}}{\partial {\bf x_{1}}}+{\bf v_{2}}\cdot
\frac{\partial g_{12}}{\partial {\bf x_{2}}}+n_{0}\int d{\bf
X_{3}}f_{1}({\bf X_3})\left( {\bf a_{13}}\cdot \frac{\partial
g_{12}}{\partial {\bf v_1}}+ {\bf a_{23}}\cdot\frac{\partial
g_{12}}{\partial {\bf v_2}}\right ) \nonumber\\
 & &=-\left ({\bf a_{12}}\cdot\frac{\partial}{\partial
{\bf v_{1}}}+{\bf a_{21}}\cdot\frac{\partial}{\partial
{\bf v_{2}}}\right )\left[f_{1}({\bf X_1})f_{1}({\bf X_2})+g_{12}\right]\nonumber\\
& &-n_{0}\int d{\bf X_{3}}\left [{\bf a_{13}}\cdot\frac{\partial
f_{1}({\bf X_1})}{\partial {\bf v_{1}}}g_{23}+{\bf
a_{23}}\cdot\frac{\partial f_{1}({\bf X_2})}{\partial {\bf
v_{2}}}g_{13}\right ]\label{v2}\end{eqnarray} where ${\bf X}={\bf (x,v)}$, $f_{1}$ is the single particle distribution, $g_{12}$ is the pair correlation function
and
$${\bf a_{ij}}=-\frac{1}{m}\frac{\partial}{\partial {\bf x_i}}{\phi_{ij}}$$
where $${\phi_{ij}}=-\frac{Gm^2}{|\bf{x_{i}}- x_{j}|}$$

The terms $a_{ij}$  denote acceleration of the i th particle due
to force exerted by the j th particle and $\phi_{ij}$ is the
energy of interaction between them. The three particle correlation function is ignored which is justified for large but finite values of N.

The system is considered to be in thermal equilibrium. Hence, the first terms of eqs. ({\ref{v1}}) and ({\ref{v2}}) are ignored. The pair correlation
function $g_{12}$ is written as
\begin{eqnarray}
 g_{12}({\bf {X_1,X_2}})&=&f_{1}({\bf X_1})f_{1}({\bf X_2})\chi_{12}({\bf x_1,x_2})
\label{v3}\end{eqnarray}

and the single particle distribution functions $f_{1}({\bf X_1})$
and $ f_{1}({\bf X_2})$ are functions of both position and
velocity. $\chi_{12}$ is a symmetric function of $\bf x_1$
and $\bf x_2$. Using eqs. ({\ref{v1}}) and ({\ref{v3}}),
\begin{eqnarray}
 {\bf v_{2}}\cdot\frac{\partial g_{12}}{\partial
{\bf x_{2}}}&=&f_{1}({\bf X_1})\chi_{12} {\bf
v_{2}}\cdot\frac{\partial
f_{1}({\bf X_2})}{\partial {\bf x_{2}}}+f_{1}({\bf X_1})f_{1}({\bf X_2}){\bf v_{2}}\cdot\frac{\partial\chi_{12}}{\partial{\bf x_{2}}}\label{v41}\\
{\bf v_{2}}\cdot\frac{\partial f_{1}({\bf
X_2})}{\partial\bf{x_{2}}} &=& - n_{0}\int d{\bf X_{3}} {\bf
a_{23}}\cdot \frac{\partial}{\partial {\bf v_{2}}} (f_{1}({\bf
X_2})f_{1}({\bf X_3})+g_{23})
\label{v4}\end{eqnarray}

and similar expressions for ${\bf v_{1}}\cdot{\partial
g_{12}}/{\partial {\bf x_{1}}}$. Using Eqs. ({\ref{v41}}) and ({\ref{v4}}) in eq. ({\ref{v2}}), we obtain:

\begin{eqnarray}
&&f_{1}({\bf X_1})f_{1}({\bf X_2})\left [{\bf
v_{1}}\cdot\frac{\partial \chi_{12}}{\partial{\bf x_{1}}}+{\bf
v_{2}}\cdot\frac{\partial \chi_{12}}{\partial{\bf x_{2}}} \right ]
\nonumber\\
& &-n_{0}\int d{\bf X_{3}}\left[f_{1}({\bf X_2})\chi_{12}{\bf a_{13}}\cdot \frac{\partial
g_{13}}{\partial {\bf v_{1}}}+f_{1}({\bf X_1})\chi_{21}{\bf a_{23}}\cdot
\frac{\partial g_{23}}{\partial {\bf v_{2}}}\right
]\nonumber\\
& &=-\left ({\bf a_{12}}\cdot\frac{\partial}{\partial {\bf
v_{1}}}+{\bf a_{21}}\cdot\frac{\partial}{\partial
{\bf v_{2}}}\right )\left (f_{1}( {\bf X_1})f_{1}({\bf X_2})+g_{12} \right )\nonumber \\
& &-n_{0}\int d{\bf X_{3}}\left [{\bf a_{13}}\cdot \frac{\partial
f_{1}({\bf X_1})}{\partial {\bf v_{1}}}g_{23}+{\bf a_{23}}\cdot
\frac{\partial f_{1}({\bf X_2})}{\partial {\bf v_{2}}}g_{13}\right
]
\label{v5}\end{eqnarray}

The single
particle distribution functions are written in the following form:
$$f_1 ({\bf X_1}) = f_M({\bf v_1})F_1({\bf x_1})  $$
where $f_M$ is a Maxwellian distribution and $F_1$ is the space part.

Inserting $g_{23}$ and $g_{13}$

\begin{eqnarray}
f_{1}({\bf X_1})f_{1}({\bf X_2})\left [{\bf
v_{1}}\cdot\frac{\partial \chi_{12}}{\partial{\bf x_{1}}}+{\bf
v_{2}}\cdot\frac{\partial \chi_{21}}{\partial{\bf x_{2}}} \right ]
=-\frac{1}{k_{B}T}\left [\frac{\partial \phi_{12}•}{\partial{\bf x_1}•}\cdot{\bf v_{1}}+\frac{\partial \phi_{12}•}{\partial{\bf x_2}•}\cdot {\bf v_{2}}\right ] f_{1}({\bf X_1})f_{1}({\bf
X_2})(1+\chi_{12})
\nonumber \\
+ f_{1}({\bf X_1})f_{1}({\bf X_2)}\frac{n_{0}m}{k_{B}T}\int d{\bf {X_{3}}}\left [{\bf {a_{13}}}\cdot {\bf {v_{1}}} f_{1}({\bf X_3}) \chi_{23}+{\bf a_{23}}\cdot
{\bf {v_{2}}} f_{1}({\bf X_3}) \chi_{13}\right]
-
\nonumber\\
 f_{1}({\bf X_1})f_{1}({\bf X_2)}\frac{n_{0}m}{k_{B}T}\int d{\bf {X_{3}}}\left [{\bf {a_{13}}}\cdot {\bf {v_{1}}} f_{1}({\bf X_3}) \chi_{13}\chi_{12}+{\bf a_{23}}\cdot
{\bf {v_{2}}} f_{1}({\bf X_3}) \chi_{23}\chi_{21}\right]\nonumber\\
\label{v6}\end{eqnarray}

We can write,
\begin{eqnarray}
f_{1}({\bf X_1})f_{1}({\bf X_2})\textbf{A}\cdot{\bf
v_{1}}
+
f_{1}({\bf X_1})f_{1}({\bf X_2})\textbf{B}
\cdot{\bf
v_{2}}=0
\label{v61}\end{eqnarray}
where
\begin{eqnarray}
\textbf{A}=\frac{\partial \chi_{12}}{\partial{\bf x_{1}}}
+\frac{1}{k_{B}T}\frac{\partial \phi_{12}•}{\partial{\bf x_1}•}(1+\chi_{12})
+ \frac{n_{0}}{k_{B}T}\int d{\bf {X_{3}}}\frac{\partial{\phi_{13}}}{{\partial {\bf x_1}}} f_{1}({\bf X_3}) \chi_{23} -\frac{n_{0}\chi_{12}}{k_{B}T}\int d{\bf {X_{3}}}\frac{\partial{\phi_{13}}}{{\partial {\bf x_1}}} f_{1}({\bf X_3}) \chi_{13}\nonumber\\
\label{v7}\end{eqnarray}
\begin{eqnarray}
\textbf{B}=\frac{\partial \chi_{21}}{\partial{\bf x_{2}}}
+\frac{1}{k_{B}T}\frac{\partial \phi_{21}•}{\partial{\bf x_2}•}(1+\chi_{21})
+ \frac{n_{0}}{k_{B}T}\int d{\bf {X_{3}}}\frac{\partial{\phi_{23}}}{{\partial {\bf x_2}}} f_{1}({\bf X_3}) \chi_{13} -\frac{n_{0}\chi_{21}}{k_{B}T}\int d{\bf {X_{3}}}\frac{\partial{\phi_{23}}}{{\partial {\bf x_2}}} f_{1}({\bf X_3}) \chi_{23}\nonumber\\
\label{v7}\end{eqnarray}

For arbitrary and linearly independent  ${\bf {v_{1}}}$ and ${\bf {v_{2}}}$, \textbf{A} and \textbf{B} both vanish.
Therefore,
\begin{eqnarray}
&&\frac{\partial \chi_{12}}{\partial{\bf x_{1}}}
+\frac{1}{k_{B}T}\frac{\partial \phi_{12}•}{\partial{\bf x_1}•}+\frac{1}{k_{B}T}\frac{\partial \phi_{12}•}{\partial{\bf x_1}•}\chi_{12}
+ \frac{n_{0}}{k_{B}T}\int d{\bf {X_{3}}}\frac{\partial{\phi_{13}}}{{\partial {\bf x_1}}} f_{1}({\bf X_3}) \chi_{23}\nonumber\\
 & & -\frac{n_{0}\chi_{12}}{k_{B}T}\int d{\bf {X_{3}}}\frac{\partial{\phi_{13}}}{{\partial {\bf x_1}}} f_{1}({\bf X_3}) \chi_{13}=0\label{v71}\end{eqnarray}
This is the equation of the pair correlation function derived from the BBGKY hierarchy. In the weak correlation limit ($\chi\ll 1$), we can drop the third and last term of the equation to obtain
\begin{eqnarray}
\frac{\partial \chi_{12}}{\partial{\bf x_{1}}}
+\frac{1}{k_{B}T}\frac{\partial \phi_{12}•}{\partial{\bf x_1}•}
+ \frac{n_{0}}{k_{B}T}\int d{\bf {X_{3}}}\frac{\partial{\phi_{13}}}{{\partial {\bf x_1}}} f_{1}({\bf X_3}) \chi_{23}=0\nonumber\\
\label{v7}\end{eqnarray}

Hence,
\begin{eqnarray}
\chi_{12}
=-\frac{1}{k_{B}T}\phi_{12}
- \frac{n_{0}}{k_{B}T}\int d{\bf {X_{3}}}\phi_{13} f_{1}({\bf X_3}) \chi_{23}
\label{v8}\end{eqnarray}
Performing the velocity integral
\begin{eqnarray}
\chi_{12} =-\frac{1}{k_{B}T}\phi_{12} - \frac{1}{k_{B}T}\int
d{\bf {r_{3}}}\phi_{13} n_{1}({\bf r_3}) \chi_{23}
\label{v9}\end{eqnarray}
The pair correlation  equation is obtained in the weak coupling limit.

\section{Results and Discussions}
In the previous section, we have derived the equation of pair correlation function  of a weakly coupled self gravitating system from BBGKY hierarchy. 

In order to obtain the equation we have assumed that the pair correlation function is the product of the single particle distribution functions and a general function of positions of the  pairing particles. This general function arises due to correlation.  

The structure of this function depends on the nature of the system. For homogenous cases, it is the sole function of $\mid \mathbf{x_{1}}-\mathbf{x_{2}}\mid$. 
For inhomogeneous cases, it can not be the sole function of $\mid \mathbf{x_{1}}-\mathbf{x_{2}}\mid$ since it does not reflect where the pair is placed in the system which is now an  important factor. Therefore, it must be a symmetric function of $\mathbf{x_{1}}$ and $\mathbf{x_{2}}$ and  should also reflect the inhomogeneity of the system.

We must choose the form of the pair correlation function in such a way that after inserting it in BBGKY equations we should obtain an equation which is applicable to both homogeneous and inhomogeneous cases. 

We may calculate the pair correlation function for homogeneous cases to check how does the equation work. 
Although self gravitating system is intrinsically  inhomogeneous, we may consider the system to be locally flat if the distance between the pair particles is much less than the size of the system. Moreover, the homogeneous cases are relatively easy to tackle mathematically. Therefore, we apply eq.(\ref{v9}) to calculate the pair correlation function in the homogeneous case. Using the following transformations 
$$\chi_{ij}=\int d\mathbf{k} \chi(\mathbf{k})e^{i\mathbf{k}\cdot(\mathbf{x_{i}}-\mathbf{x_{j}})}$$
$$\phi_{ij}=-\frac{Gm^2}{2\pi^2}\int\frac{d\mathbf{k}}{k^{2}}e^{i\mathbf{k}\cdot(\mathbf{x_{i}}-\mathbf{x_{j}})}$$ in eq.(\ref{v9}) we obtain

\begin{eqnarray}
\chi(\mathbf{k})=\frac{1}{2\pi^2k_{B}T}\frac{Gm^{2}}{k^{2}-k_{j}^{2}}
\label{v10}\end{eqnarray}
where $k_{j}^2=4\pi n_{0}Gm^{2}/k_{B}T$. 
Therefore, the pair correlation function with collective effects is obtained as
\begin{eqnarray}
\chi_{ij}=\frac{1}{k_{B}T}\frac{Gm^{2}}{r}\cos{(k_{j}r)}
\label{v11}\end{eqnarray}
where $ r=\mid \mathbf{x_{i}}-\mathbf{x_{j}}\mid$
Hence, the potential of interaction is
$$-\frac{Gm^{2}}{r}\cos{(k_{j}r)}$$.
which is identical to what has been shown by Chavanis \cite{kn:ph4}.  We could neglect the collective effects by switching off the last term of eq.(\ref{v9}) to obtain
\begin{eqnarray}
\chi_{ij}=\frac{1}{k_{B}T}\frac{Gm^{2}}{r}
\label{v12}\end{eqnarray}

It is interesting to note that in the limit $k_{j}r<<1$ the result with collective effects (eq.(\ref{v11})) tends to become identical to the result without collective effects (eq.(\ref{v12})). It has a serious implication on self gravitating system.  Since, as described by Binney 
and Treamine\cite{kn:sc}, equal octaves in impact
parameter contribute equally to gravitational scattering, most of the contributions of the two body interactions are local in nature i.e. the distance between the pairing particles is much less than the Jean's length. 
Therefore, under such circumstances, we could effectively use the result without collective effect in place of the result with collective effects. It may be the reason that the theory of Chandrasekhar\cite{kn:ch1,kn:ch2,kn:ch3} which did not consider the collective effects,  gives a reasonably good description of the self gravitating system. Obviously there is  scope to incorporate the collective effects in the system to make the calculation more rigorous. 
We may expect the above discussion to be true even for the inhomogeneous system. After all the inhomogeneous system appears to be homogeneous for the shorter length scale and the underlying  physics of collective phenomenon does remain same in both the cases. If we increase the distance between the pairing particles the collective effects should reflect in the expression of the pair correlation function. Intuitively, some symmetric function of positions, other than the function of $\mid \mathbf{x_{i}}-\mathbf{x_{j}}\mid$, may appear in the expression. For example, it may be a function of $(\mathbf{x_{1}}+\mathbf{x_{2}})$ which is  the simplest function  among the functions which are symmetric in positions of the particles. 
Finally, consideration of strong  inhomogeneity needs numerical techniques which is beyond the scope of this work. 

\section{Conclusion}

We have obtained an equation of pair correlation function from the first two members of BBGKY hierarchy for a weakly interacting inhomogeneous self gravitating system confined by a finite spherical container in thermal equilibrium. Application of artificial container is not uncommon in the literature \cite{kn:va,kn:dl,kn:tp,kn:jj} of many body self gravitating system. In our case, presence of artificial spherical container is necessary to prohibit the particles to escape from the system and achieve the thermal equilibrium state.

A real self gravitating system never reach thermal equilibrium due to the evaporation of particles from the system.  However,  the central part of a star cluster that is a few or more times older than its central relaxation time will be close to the thermal equilibrium state \cite{kn:sc} and in that particular situation  we may ignore the high-energy regions close to the escape energy and those cases in which the system is unstable to gravothermal collapse. Hence, this formalism may be useful to explore the thermodynamic properties of such system.

The pair correlation function $\chi_{12}$ is a symmetric function of $\bf{x_{1}}$ and $\bf{x_{2}}$. However, in contrast to the homogeneous and locally flat cases, it is not the sole function of $|\bf{x_{1}}- x_{2}|$. Consequently, we are not bound to consider our system to be locally flat. In addition to that,
the presence of third term of eq.(\ref{v7}) confirms that this formalism also accounts for the collective effects.

\subsection{Acknowledgement }
I would like to 
acknowledge helpful discussions with Prof. S. Tremaine.

\end{document}